\documentclass[10pt,conference]{IEEEtran}

\usepackage{cite}
\usepackage{amsmath,amssymb,amsfonts}
\usepackage{algorithmic}
\usepackage{graphicx}
\usepackage{textcomp}
\usepackage{xcolor}
\usepackage{todonotes}
\usepackage{url}
\usepackage{subfig}

\newcommand{\eventually}[0]{\Diamond}
\newcommand{\globally}[0]{\Box}

\newcommand{\signal}[0]{\textbf{s}}

\def\BibTeX{{\rm B\kern-.05em{\sc i\kern-.025em b}\kern-.08em
    T\kern-.1667em\lower.7ex\hbox{E}\kern-.125emX}}
    
\begin{document}



\title{Runtime Resolution of Feature Interactions  through Adaptive Requirement Weakening}

\author{\IEEEauthorblockN{Simon Chu}
\IEEEauthorblockA{\textit{Carnegie Mellon University} \\
Pittsburgh, PA USA \\
cchu2@andrew.cmu.edu}
\and
\IEEEauthorblockN{Emma Shedden}
\IEEEauthorblockA{\textit{University of Michigan} \\
Ann Arbor, MI USA\\
emshedde@umich.edu}
\and
\IEEEauthorblockN{Changjian Zhang, R\^omulo Meira-G\'oes}
\IEEEauthorblockA{\textit{Carnegie Mellon University} \\
Pittsburgh, PA USA \\
\{changjiz, rmeirago\}@andrew.cmu.edu}
\and
\IEEEauthorblockN{Gabriel A. Moreno}
\IEEEauthorblockA{\textit{Software Engineering Institute, Carnegie Mellon University} \\
Pittsburgh, PA USA \\
gmoreno@sei.cmu.edu}
\and
\IEEEauthorblockN{David Garlan, Eunsuk Kang}
\IEEEauthorblockA{\textit{Carnegie Mellon University} \\
Pittsburgh, PA USA \\
\{dg4d, eunsukk\}@andrew.cmu.edu}
}

\maketitle

\thispagestyle{plain}
\pagestyle{plain}

\begin{abstract}
The \emph{feature interaction problem} occurs when two or more independently developed components interact with each other in unanticipated ways, resulting in undesirable system behaviors. Feature interaction problems remain a challenge for emerging domains in cyber-physical systems (CPS), such as the Internet of Things and autonomous drones. Existing techniques for resolving feature interactions take a ``winner-takes-all'' approach, where one out of the conflicting features is selected as the most desirable one, and the rest are disabled. However, when multiple of the conflicting features fulfill important system requirements, being forced to select one of them can result in an undesirable system outcome. In this paper, we propose a new resolution approach that allows all of the conflicting features to continue to partially fulfill their requirements during the resolution process. In particular, our approach leverages the idea of \emph{adaptive requirement weakening}, which involves one or more features temporarily \emph{weakening} their level of performance in order to co-exist with the other features in a consistent manner. Given feature requirements specified in Signal Temporal Logic (STL), we propose an automated method and a runtime architecture for automatically weakening the requirements to resolve a conflict. We demonstrate our approach through case studies on feature interactions in autonomous drones. 
\end{abstract}


\section{Introduction}

Modern software systems are often constructed by composing a set of independently developed components or \emph{features}, each of which is designed to achieve a particular objective or a \emph{requirement}. For instance, a typical automotive system contains an array of software features that are designed to ensure vehicle safety under different circumstances, such as emergency braking and lane-keeping assist. 

Sometimes, these features can interact with each other in unanticipated ways, resulting in undesirable system behavior. This type of problem, also called the \emph{feature interactions problem}, has been long studied by the software engineering community\cite{fi-intro,nuseibeh-security,zave-fi}, but remains a major challenge, especially in emerging domains such as the Internet of Things and autonomous systems~\cite{fi-iot,alma08,metzger04,atlee-fse17}. There are two main aspects to the feature interactions problem: (1) \emph{detection} of a possible conflict between features and (2) its \emph{resolution}. In this paper, we focus on the resolution of  feature interactions at \emph{run-time}.

Existing approaches to resolution leverage some notion of what it means for one feature to be more \emph{desirable} than others; then when a conflict arises, the most desirable out of the conflicting features is selected as the one that is ultimately executed by the system (and actions from the rest are discarded). One such common notion is based on user-defined \emph{priorities} or \emph{precedent list}~\cite{lafortune-resolution,atlee-resolution,atlee-ordering,leung-ordering}. Other recent approaches include \emph{variable-specific} resolution (where a conflict between features that modify the same variable is resolved by selecting the action that is considered safest~\cite{atlee-fse14,atlee-fse17}) and \emph{property-based} resolution (where the feature that is most likely to satisfy a given system requirement is selected~\cite{rv18}). 

These ``winner-takes-all'' approaches, however, share one major drawback: When multiple of the conflicting features fulfill important system requirements, being forced to select only one of them leads to an outcome that is undesirable from the developer's perspective. For example, when a pair of conflicting features in a vehicle are both designed to perform critical safety functions (e.g., maneuvering around an  obstacle while staying within the lane), discarding one or the other might bring the system into an unsafe state in either case. 

To overcome this challenge, we propose a new resolution approach that allows the conflicting features to continue to (partially) fulfill their requirements during the resolution process. The key idea behind this approach is that of \emph{adaptive requirement weakening}: When a feature conflicts with another, it may be acceptable to temporarily ``compromise'' the level of its functionality, by weakening the requirement that it is designed to achieve. Consider a pair of features, $F_1$ and $F_2$, that are designed to fulfill requirements $R_1$ and $R_2$, respectively. When in presence of each other, there may be  scenarios in which both requirements cannot be fulfilled simultaneously (i.e., $F_1 \oplus F_2 \not\models R_1 \land R_2$). To resolve this conflict, instead of  disabling $F_1$ or $F_2$, our approach involves weakening one or more of the given feature requirements (e.g., from $R_1$ to $R'_1$) such that the features are able to function and co-exist with each other in a consistent manner (e.g., $F_1 \oplus F_2 \models R'_1 \land R'_2$). In addition, we say that  our approach is \emph{adaptive}, since the degree to which one or more requirements are weakened depends on the particular environmental context in which a conflict arises. 

In this paper, we present a realization of this approach in a class of systems called \emph{cyber-physical systems} (CPS), where software features are used to monitor and control one or more physical entities in the environment~\cite{parnas-four-variable}. The requirements of individual features are specified using a notation called \emph{Signal Temporal Logic (STL)}~\cite{stl}, which is particularly well-suited for describing continuous-domain and time-sensitive behaviors of CPS. Based on the semantics of STL~\cite{stl}, we provide (1) a formal definition of what it means to \emph{weaken} a requirement, (2) a method for automatically transforming a pair of conflicting requirements, $R_1$ and $R_2$, into weakened, consistent versions, $R'_1$ and $R'_2$, and (3) a runtime architecture that leverages this method to dynamically modify the behavior of the components and resolves the conflict. 

Weakening the requirement of a feature, however, involves degrading the level of its functionality and can reduce the overall utility of the system. Thus, an ideal method would weaken the requirements no more than by a \emph{minimal} degree that is sufficient to resolve the given conflict. Finding such \emph{minimal weakenings}, however, is a challenging problem, since in general, the space of weakening candidates for a given requirement $R$ can be enormous, if not infinite. In particular, we demonstrate how this problem can be formulated as an instance of \emph{mixed-integer linear programming (MILP)}~\cite{Papadimitriou82}, where the goal is to synthesize weakened requirements that no longer conflict with each other while being as close to the original requirements as possible.

We have built a prototype implementation of our runtime resolution approach on top of PX4 Autopilot~\cite{px4}, an open-source autopilot software used in consumer and industrial drones. We have evaluated our approach using four different (possibly conflicting) features in an autonomous drone under a wide range of simulated scenarios. Our evaluation shows that our weakening-based approach is  effective at resolving conflicts while allowing the features to continue to satisfy the weakened versions of their requirements.

The contributions of this paper are as follows:
\begin{itemize}
    \item A theoretical foundation for the resolution of feature interactions using STL-based requirement weakening (Section~\ref{sec:weakening})
    \item A runtime architecture that leverages requirement weakening to resolve conflicts (Section~\ref{sec:runtime-architecture}),
    \item An approach for finding minimal weakening through translation into MILP (Section~\ref{sec:milp}), and;
    \item An implementation of the weakening-based resolver (Section~\ref{sec:implementation}) and an evaluation on case studies involving autonomous drone features (Section~\ref{sec:evaluation}).
\end{itemize}
\noindent\textbf{Relevance to SEAMS}. Our  approach can be regarded as performing a type of self-adaptation, as it involves dynamically changing the behavior of components to manage conflict scenarios that are difficult to predict or resolve at the design time. As opposed to the typical control loop used in self-adaptive systems (e.g., MAPE-K~\cite{sas-roadmap}), which adapts the system asynchronously, our approach is synchronous with the control loop of the CPS, as later described in Section~\ref{sec:runtime-architecture}, Fig.~\ref{fig:resolution-overview}.


\section{Motivating Example}\label{sec:motivating-example}



Consider an autonomous organ delivery drone attempting to deliver an organ from a donor in \textit{Hospital B} to a recipient in \textit{Hospital A}, inspired by the example from~\cite{Maia2019}. The drone contains two features: (1) the \textit{delivery path planner}, which ensures the organ is delivered to the receiver's hospital in the most efficient manner possible, and (2) the \textit{safe landing enforcer}, which ensures that the battery on the drone has enough charge to safely make it to the nearest land. Under normal conditions, the \textit{delivery path planner} will always be active, generating an action ($a_{deliver}$) to move the drone to its destination, while the \textit{safe landing enforcer} remains off by default unless triggered upon sensing a low battery level. 

\begin{figure}[!t]
  \centering
  \includegraphics[width=0.75\linewidth]{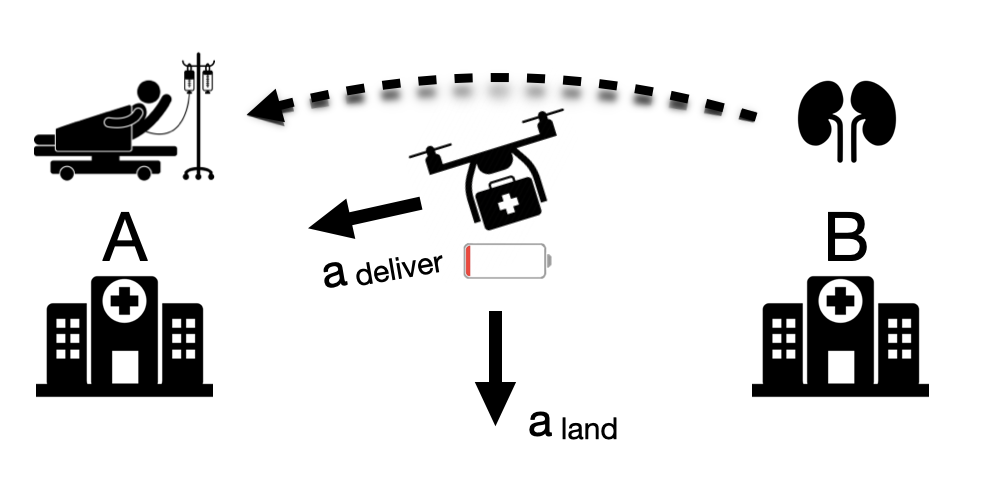}
  \caption{\small{A possible conflict in an organ delivery drone.}}
  \vspace{-10pt}
  \label{fig:organ-delivery}
\end{figure}

Imagine a scenario where the drone encounters unexpected turbulence mid-flight, causing the battery to discharge much faster. When the battery dips below a certain threshold, the safe landing enforcer is activated, which then generates a safe landing action ($a_{land}$) to direct the drone to the nearest land. Since the delivery path planner is unaware of the landing enforcer, it continues to generate $a_{deliver}$, which results in a conflict between the two features, as depicted in Figure~\ref{fig:organ-delivery}.


\paragraph{Existing methods} 
One possible approach to resolving this conflict is to leverage a user-defined list of priorities among the features. The drone operator, for example, may designate the safe landing feature as having the highest priority, and the flight controller could be programmed to disregard the actions from all other features, including the delivery path planner. This approach to resolution results in system behaviors where the requirement of the highest-ranked feature (i.e., safe landing) is satisfied while the others are disregarded; in this example, choosing to land the drone may result in the organ failing to be delivered before its expiry time. 

This type of “winner-takes-all” approach, however, may not be suitable in situations where such a strict ordering among features does not exist, or all of the conflicting features play a critical role in maintaining the system's safety and performance. For instance, while landing the drone safely before running out of battery is certainly important,  giving up an organ in the process is also a highly undesirable outcome, as the life of a patient may depend on its timely delivery.



\paragraph{Proposed method} 
Our  approach, in comparison, attempts to resolve the conflict in a way that satisfies the requirements of both  conflicting features. The key idea is that for certain types of requirements in CPS, it may be acceptable to temporarily compromise the degree to which the system satisfies them; i.e., satisfy a weaker version of an original requirement. This notion of requirement satisfaction, also termed \emph{satisficing}~\cite{simon1956rational}, can enable a resolution approach that involves relaxing some of the requirements but without entirely giving up on any one of them.











For example, suppose that the requirements for the safe landing and delivery planning features are as follows:
\begin{quote}
$R_{land}$: \emph{If the battery threshold falls below 10\%, the drone should land on the nearest land.}\\
$R_{deliver}$: \emph{The drone should fly at a fast-enough speed to reach the destination before the delivery time.}
\end{quote}
During resolution, one or both of these requirements can be weakened. For example, the requirement for the safe landing feature may be weakened by lowering the threshold that triggers the drone to find a landing spot (e.g., from 10\% to 5\%). With the new weakened requirement ($R_{land}$ $\rightarrow$ $R'_{land}$), the behavior of the two features  becomes consistent again (i.e., $R'_{land} \land R_{delivery}$ is satisfiable). Intuitively, this amounts to delaying the safe landing feature in order to allow the drone to complete its organ delivery mission. As a trade-off, the safe landing feature has compromised the level of safety that it originally promised, since there is now an increased risk that the drone may run out of battery before landing. 

Alternatively, the conflict could be resolved by weakening both requirements. For example, $R_{delivery}$ may be weakened by reducing the speed of the drone, to allow the battery to be depleted at a slower rate than at the original speed. This would cause a delay in organ delivery, but it would also enable the drone to complete its delivery while weakening the battery threshold from 10\% to 8\% only. In either case, this weakening-based approach results in an arguably more desirable system outcome than the priority-based method, since both the requirements of the features can be satisfied (at the cost of temporarily compromising their optimal functionality). 

\paragraph{Challenges} In general, there may be a large number of  ways to resolve a conflict using this approach, weakening one or more requirements by different amounts. Since weakening involves degrading the functionality of the features, an ideal resolution process would involve weakening the requirements no more than needed. At the same time, the system operator may also wish to place a harder constraint on the maximum amount by which a requirement can be weakened (e.g., for $R_{land}$, the threshold cannot be set below 5\%, since that might compromise the safety of the battery beyond what is acceptable). We later show (1) how this type of weakening can be formally realized over STL and (2) an approach that uses a MILP solver to generate minimal weakenings.

\section{Preliminaries}\label{sec:background}

\paragraph{Signals}
In our approach, the behavior of CPS is modeled by real-valued continuous-time \emph{signals}.
Formally, a signal $s$ is a function $\signal{}:T\rightarrow D$ mapping from a time  domain, $T\subseteq \mathbb{R}_{\geq 0}$, to a tuple of $k$ real numbers, $D\subseteq \mathbb{R}^k$.
Intuitively, the value of a signal $\signal{}(t) = (v_1,\dots, v_k)$ represents different state variables of the system at time $t$; (e.g., $v_1$ might represent the altitude of the drone).

\paragraph{Signal Temporal Logic (STL)}
STL extends linear temporal logic (LTL) \cite{ltl} for specifying the time-varying behavior of a system in terms of signals. The basic unit of a formula in STL is a signal predicate in the form of $f(\signal{}(t)) > 0$, where $f$ is a function from $D$ to $\mathbb{R}$; i.e., the predicate is true if and only if $f(\signal{}(t))$ is greater than zero. Then, the syntax of an STL formula $\varphi$ is defined as:
\begin{equation*}
    \varphi := \top \mid f(\signal{}(t)) > 0 \mid \neg \varphi \mid \varphi_1\wedge \varphi_2\mid \varphi_1 \mathcal{U}_{[a,b]}\varphi_2
\end{equation*}
where $\top$ is a Boolean $true$ constant, $a,b\in \mathbb{R}$  and $a<b$. The \emph{until} operator $\varphi_1 \mathcal{U}_{[a,b]}\varphi_2$ means that $\varphi_1$ must hold until $\varphi_2$ becomes true within a time interval $[a,b]$.
The until operator can be used to define two other important temporal operators: \emph{eventually} ($\Diamond_{[a,b]}\varphi := True\ \mathcal{U}_{[a,b]}\varphi$) and \emph{always} ($\Box_{[a,b]}\varphi := \neg\Diamond_{[a,b]}\neg \varphi$). 

\paragraph{Robustness}
Typically, the semantics of temporal logic such as LTL is based on a \emph{binary} notion of formula satisfaction (i.e., formula $\varphi$ is either satisfied or violated by the system). Due to its signal-based nature, STL also supports a \emph{quantitative} notion of satisfaction, which allows reasoning about how ``close" or ``far'' the system is from satisfying or violating a property. This quantitative measure is also called the \emph{robustness} of satisfaction. 

Informally, the robustness of signal $\signal{}$ with respect to formula $\varphi$ at time $t$, denoted by $\rho(\varphi,\signal{},t)$, represents the smallest difference between the actual signal value and the threshold at which the system violates $\varphi$.
For example, if the property $\varphi$ says that ``the drone should maintain an altitude of at least 5.0 meters," then $\rho(\varphi,\signal{},t)$ represents how close to 5.0 meters the drone maintains its altitude.
Formally, robustness is defined over STL formulas as follows:
\begin{align}
    \rho(f(\signal{}(t)) > 0,\signal{},t) & \equiv f(\signal{}(t))\nonumber\\
    \rho(\neg \varphi, \signal{}, t) &\equiv -\rho(\varphi,\signal{},t) \nonumber \\
    \rho(\varphi_1\wedge \varphi_2,s,t) &\equiv \min\{\rho(\varphi_1,\signal{},t),\rho(\varphi_2,\signal{},t)\}\nonumber\\
    \rho(\Diamond_{[a,b]}\varphi,\signal{},t) &\equiv \sup_{t_1\in[t+a,t+b]} \rho(\varphi,\signal{},t_1) \nonumber\\
    \rho(\Box_{[a,b]}\varphi,\signal{},t) &\equiv \inf_{t_1\in[t+a,t+b]} \rho(\varphi,\signal{},t_1)\nonumber 
\end{align}
where $\inf_{x \in X} f(x)$ is the greatest lower bound of some function $f:X\rightarrow \mathbb{R}$ (and $\sup$ the least upper bound).
The robustness of satisfying predicate $f(\signal{}(t)) > 0$ captures how close signal $\signal{}$ at time $t$ is above or below zero.
For example, consider formula $\varphi \equiv alt(t) - 5 > 0$, capturing the property that ``the drone altitude is at least 5.0 meters."
If, at time $t$, the altitude signal is $alt(t) = 10$ meters (i.e., $5.0$ meters above the required altitude), $\rho(\varphi,\signal{},t)$ is computed as  $5$.

On the other hand, robustness $\rho(\Box\varphi,a,t)$ describes the point at which the system is furthest away from satisfying $\varphi$. For instance, consider property $\phi$ $\equiv$ $\Box_{[0,2]}(alt(t) - 5 > 0)$, which says that the drone must maintain a minimum altitude of 5.0 meters for interval $t=[0, 2]$. Suppose that the system evolves to generate signal $\signal{}$ with the altitude of 6.0, 3.0, 5.5 meters at $t=0, 1, 2$, respectively; then, $\rho(\phi, \signal{},0)$ = $\rho((alt(t) - 5 > 0), \signal{},1)$ = -2.0 (i.e., the system \emph{violates} $\phi$ by the robustness value of 2.0).





\section{Requirement Weakening}
\label{sec:weakening}
We present an extension to STL to support the systematic weakening of requirements in STL. It enables us to specify the maximum extent of weakening allowed for a requirement and quantitatively measure the degree of weakening, which later plays an important role to ensure that requirements are weakened no more than needed to resolve a conflict. The extension is built upon the observation that the robustness value of an instantiated STL formula can be increased by changing the atomic proposition, and shortening or enlarging the time interval of the temporal operators in STL. This increase in robustness quantification leads to a less restrictive requirement, which is easier to satisfy than the original one.

\subsection{weakSTL: STL with Weakening}
We propose \emph{weakSTL}, an extension to STL with weakening semantics. \emph{weakSTL} introduces additional parameters to atomic propositions and temporal operators to indicate the range of signal values and time intervals allowed for weakening. The syntax of a \emph{weakSTL} formula is defined as:
\begin{align*}
    \varphi ~:=~    & \top                          ~\mid~
                    f_p(\signal{}(t)) > 0           ~\mid~
                    \neg \varphi                    ~\mid~
                    \varphi \land \psi              ~\mid~
                    \varphi \lor \psi               ~\mid~ \\
                    & \globally{}_{I, p, q} \varphi ~\mid~
                    \eventually{}_{I, p, q} \varphi ~\mid~
                    \varphi ~\mathcal{U}_I~ \psi
\end{align*}
where $I$ represents the original time interval $[a, b]$, and $p, q$ represent the weakening parameters that define the maximum range for the signal values and time intervals to be weakened. Note that, our definition does not allow weakening parameters over the \emph{Until} operator ($\varphi ~\mathcal{U}_{[a, b]}~\psi$) because when $\varphi$ and $\psi$ are non-trivial formulas, changing the time interval ($[a, b]$) has a mixed-effect over the difficulty of satisfaction; i.e., it cannot guarantee the resulting formula will be  weaker. For example, increasing $b$ allows $\psi$ to occur later (which makes it easier to satisfy) but  requires $\varphi$ to hold longer (which makes it harder).


\subsubsection{Semantics} Formally, a \emph{weakSTL} formula $\varphi$ defines a set of STL formulas that are weaker variants of the original STL formula. 
We define the translation rules $M: weakSTL \rightarrow \mathbb{P}(STL)$ as follows:

{\small
\begin{align*}
    M(\top) & = \{\top\} \\
    M(f_p(\signal{}(t)) > 0) & = \{f(\signal{}(t)) + i > 0 ~|~ 0 \le i \le p\} \\
    M(\neg \varphi) & = \{\neg \varphi' ~|~ \varphi' \in M_s(\varphi)\} \\
    M(\varphi \land \psi) & = \{\varphi' \land \psi' ~|~ \varphi' \in M(\varphi), \psi' \in M(\psi)\} \\
    M(\varphi \lor \psi) & = \{\varphi' \lor \psi' ~|~ \varphi' \in M(\varphi), \psi' \in M(\psi)\} \\
    M(\globally{}_{[a, b], p, q} ~\varphi) & = \{\globally{}_{[a+i, b-j]}~\varphi' ~|~ \\  & \quad \quad 0 \le i \le p, 0 \le j \le q, \varphi' \in M(\varphi)\} \\
    M(\eventually{}_{[a, b], p, q} ~\varphi) & = \{\eventually{}_{[a-i, b+j]}~\varphi' ~|~ \\ & \quad \quad 0 \le i \le p, 0 \le j \le q, \varphi' \in M(\varphi)\} \\
    M(\varphi~\mathcal{U}_{[a, b]}~\psi) & = \{\varphi'~\mathcal{U}_{[a, b]}~\psi' ~|~ \varphi' \in M(\varphi), \psi' \in M(\psi) \} \\
\end{align*}
}%
For example, consider $\varphi \equiv f_{p}(\signal{}(t)) > 0$ for some given value $p$, where $f(\signal{}(t)) \equiv alt(t) - 5$. Then, $M(\varphi)$ represents the set of all STL formulas of the form $alt(t) - 5 + i > 0$ for $0 \leq i \leq p$; in other words, $M(\varphi)$ represents weaker variants of $\varphi$ that requires the drone to maintain an altitude higher than ($5 - i$) only instead of $5$ meters in the original $\varphi$.

Given $\varphi \equiv \globally{} \phi$, $M(\varphi)$ represents the set of formulas where the system is required to maintain $\phi$ throughout a \emph{narrower} interval than the one specified by the original formula $\varphi$. Similarly, for $\varphi \equiv \eventually{} \phi$, $M(\varphi)$ relaxes this requirement to allow the system to satisfy $\phi$ just once during a larger window than in $\varphi$, which is easier to fulfill.

Note that weakening formulas of the form $\neg \varphi$ involves \emph{strengthening} $\varphi$. To achieve this, we also introduce $M_s$, which defines the translation rules for strengthening an STL formula. In short, $M_s$ is similar to $M$ but  inverses the weakening computations in $M$:

{\small
\begin{align*}
    M_s(f_p(\signal{}(t)) > 0) & = \{f(\signal{}(t)) - i > 0 ~|~ 0 \le i \le p\} \\
    M_s(\neg \varphi) & = \{\neg \varphi' ~|~ \varphi' \in M(\varphi)\} \\
    M_s(\globally{}_{[a, b], p, q} ~\varphi) & = \{\globally{}_{[a-i, b+j]}~\varphi' ~|~ \\ & \quad \quad 0 \le i \le p, 0 \le j \le q, \varphi' \in M_s(\varphi)\} \\
    M_s(\eventually{}_{[a, b], p, q} ~\varphi) & = \{\eventually{}_{[a+i, b-j]}~\varphi' ~|~ \\ & \quad \quad 0 \le i \le p, 0 \le j \le q, \varphi' \in M_s(\varphi)\}\\
\end{align*}
}%

\subsubsection{Instantiation} By assigning concrete weakening values to each of the weakened operators and propositions, we can instantiate an STL formula from a \emph{weakSTL} formula. Formally, we state this as $\varphi_\theta = \kappa(\varphi, \theta)$, where $\kappa$ is the \emph{instantiation function}, $\varphi$ is a \emph{weakSTL} formula, $\varphi_\theta$ is an STL formula, and $\theta$ assigns concrete weakening values within the range given by the weakening parameters (i.e., $p$, $q$'s) in $\varphi$.

Specifically, $\theta$ is a partial function of type $weakSTL \rightarrow  \mathbb{Z}^k$, where $k$ equals the total number of weakening parameters in $\varphi$. Given $\varphi$, $\theta$ maps each sub-expression of $\varphi$ to concrete values that are  used to determine how much that sub-expression is weakened. Formally, we have:

{\small
\begin{align}
    \theta(\top) & = () \nonumber\\
    \theta(f_p(\signal{}(t)) > 0) & = (x), ~ 0 \leq x \leq p \nonumber \\
    \theta(\neg \varphi) & = \theta(\varphi) \nonumber \\
    \theta(\varphi \land \psi) & = \theta(\varphi)^{\frown}\theta(\psi) \nonumber \\
    \theta(\varphi \lor \psi) & = \theta(\varphi)^{\frown}\theta(\psi) \nonumber \\
    \theta(\globally{}_{[a, b], p, q} ~\varphi) & = (x, y)^{\frown}\theta(\varphi), ~ 0 \leq x \leq p \land 0 \leq y \leq q \nonumber\\
    \theta(\eventually{}_{[a, b], p, q} ~\varphi) & = (x, y)^{\frown}\theta(\varphi), ~ 0 \leq x \leq p \land 0 \leq y \leq q \nonumber\\
    \theta(\varphi~\mathcal{U}_{[a, b]}~\psi) & = \theta(\varphi)^{\frown}\theta(\psi) \nonumber \\ \label{eq:weakening-parameters}
\end{align}
}%

Then, the instantiation function $\kappa$ takes a given mapping $\theta$ and generates an STL formula based on the translation rules $M$ and $M_s$, i.e., plugging in values from $\theta$ into  $i, j$'s as it walks over the sub-expressions of $\varphi$ recursively.\footnote{$s_1^{\frown}s_2$ represents the result of concatenating sequences $s_1$ and $s_2$.}


We say that a \emph{weakSTL} formula $\varphi$ is satisfiable when there exists an instantiated STL formula that is satisfiable, i.e., $$(\signal{}, t) \models \varphi \Leftrightarrow \exists \varphi_\theta \in M(\varphi) ~\bullet~ (\signal{}, t) \models \varphi_\theta$$
In addition, we define $\varphi_0$ be the instantiation without any weakening from the \emph{weakSTL} formula $\varphi$, i.e., $\theta$ is defined to be 0 for every sub-expression of $\varphi$.

Finally, we leverage the robustness concept in the original STL to measure the \emph{degree of weakening} between two instantiations of a \emph{weakSTL} formula $\varphi$, as follows: $$\Delta(\varphi_{\theta 1}, \varphi_{\theta 2}, \signal{}, t) = \rho(\varphi_{\theta 2}, \signal{}, t) - \rho(\varphi_{\theta 1}, \signal{}, t)$$ where $\varphi_{\theta 1}, \varphi_{\theta 2} \in M(\varphi)$.

\subsubsection{Example}

Let us revisit the example in Section \ref{sec:background}, where $\phi \equiv \Box_{[0,2]}(alt(t) - 5 > 0)$. Consider the \emph{weakSTL} formula $\varphi \equiv \Box_{[0,2], 0, 2}(alt(t) - 5 > 0)$ (i.e., $\varphi = M(\phi)$ with $p = 0$ and $q = 2$). Then, $\varphi$ can be instantiated into two weaker versions of $\varphi$, by weakening the time interval to $t' = [0, 1]$ with $\theta = (0, 1)$ or $t' = [0, 0]$ with $\theta = (0, 2)$. 

Given signal \signal{} with an altitude of 6.0, 3.0, 5.5 meters at $t = 0, 1, 2$, respectively, the original STL requirement $\phi$ is violated, with the robustness value of -2.0. However, with $\theta = (0, 2)$, the resulting, weaker STL requirement $\varphi_{\theta}$ is satisfied, with the robustness value of $\rho(\varphi_{\theta}, \signal, 0) = 1.0$. The degree of weakening between the weaker and initial requirements is measured as $\Delta(\varphi_0, \varphi_\theta) = \Delta(\phi, \varphi_\theta) = 1.0 - (-2.0) = 3.0$.

\section{Runtime Resolution Architecture}
\label{sec:runtime-architecture}

\begin{figure}[!t]
  \centering
  \includegraphics[width=0.95\linewidth]{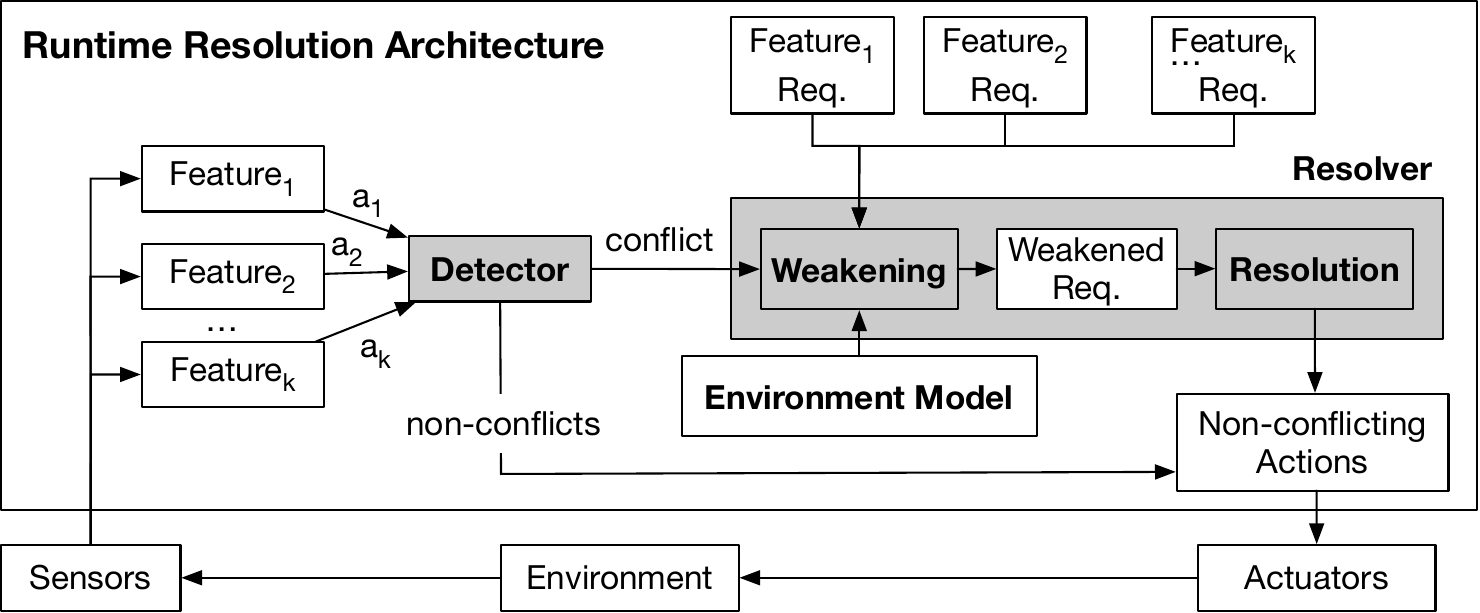}
  \caption{\small{Overview of the proposed resolution architecture.}}
  \label{fig:resolution-overview}
  \vspace{-15pt}
\end{figure}

An overview of our proposed runtime architecture for weakening-based  resolution is shown in Figure~\ref{fig:resolution-overview}. We assume that each feature in our system periodically observes the state of the environment (through one or more sensors) and generate a command to an actuator in order to influence the state. For example, the safe landing feature from our running example periodically monitors the state of the battery (which is a physical component and is thus considered part of the environment for the software system). 
If the battery level drops below a safe threshold, the landing feature generates a command to direct the drone to land on the nearest land.

There are two major components in the proposed architecture: (1) the \emph{detector}, which detects possible conflicts between the features, and (2) the \emph{resolver}, which resolves a possible conflict by generating a new set of actions that are consistent with each other. Since the focus of this paper is on resolution, for detection, we adopt the approach proposed by~\cite{atlee-fse17} and \cite{atlee-fse14}, where a pair of features are considered to be in conflict if they generate actions that modify an overlapping part of the environment (e.g., the safe landing and delivery path planning features both affect the direction of the drone's next movement and thus are in conflict). 

When triggered by the detector, the resolver takes three types of inputs: (1) a conflict, represented by a set of conflicting features, (2) a set of requirements for the conflicting features, and (3) an \emph{environment model}, which describes how the state of the environment evolves based on the action of the system (explained further below). Then, the resolver performs two steps: (1) \emph{weakening} of one or more of the given requirements, and (2) produce a set of actions of the features that adhere to the weakened requirements (and thus, non-conflicting with each other). 

In the subsequent sections, we further explain (1) how an environment model is used to predict the behavior of a system given an action and (2) how the weakening and resolution steps by the resolver are carried out using a MILP solver.





\subsection{Environment Model}


Given a set of feature requirements, $R_1,...,R_k$, the goal of the resolution is to find weaker versions of one or more of them, $R'_1,...,R'_k$, such that they are consistent (i.e., it is possible to satisfy $R'_1 \land ... \land R'_k$). In the context of STL, checking the satisfaction of weakened requirement $R'$ involves evaluating it over some signal $\signal{}$ that represents the possible future states of the system \emph{if} the component behaved according to $R'$.
In the proposed framework, the environment model plays the  role of generating such a \emph{predictive signal}.



  
More specifically, the environment model is represented as transition system $T = (\mathcal{Q}, \mathcal{A}, \mathcal{\delta}, \mathcal{Q}_i)$, where:
\begin{itemize}
    \item $\mathcal{Q} \subseteq \mathbb{R}^{k}$ is the set of environment states. Each state is a particular combination of values for signal variables, represented as a k-dimensional tuple;  $q = (v_1, ... v_k) \in \mathcal{Q}$.
    \item $\mathcal{A}$ is the set of actuator actions.
    \item $\mathcal{\delta}: \mathcal{Q} \times \mathcal{A} \rightarrow \mathcal{Q}$ is the transition function that captures how the system moves from one state to another by performing an action.
    \item $\mathcal{Q}_i$ is the set of initial states.
\end{itemize}
For example, the environment model for the drone example may capture the (x, y, z) location of the drone, its velocity, as well as the amount of remaining battery. The location of the drone changes during each transition depending on the current velocity, which, in turn, may be modified by a system action that accelerates or decelerates the drone. Similarly, the battery level also can be modeled as decreasing at a steady rate ($drain\_rate$) while the drone is in movement:
\begin{align*}
&  q' = \delta(q,a)  \\
& q'.battery\_level = q.battery\_level - drain\_rate
\end{align*}
Then, given a sequence of actions $a_1,a_2,...,a_n$ and the current state $q$, the environment model can be executed over these actions to generate a corresponding state sequence, $q_0, q_1, ... q_n$, which can then be formed into predictive signal \signal{}.

Our approach does not prescribe a particular notation for specifying an environment model, as long as it can be used to generate signals as depicted above. For our implementation, we use the MiniZinc modeling language~\cite{minizinc}, which provides declarative constraints for specifying relationships between different variables of a system. 




\subsection{Weakening-based Resolution as MILP}
\label{sec:milp}




\newtheorem{problem}{Problem}

In our approach, the weakening and resolution steps inside the resolver (Figure~\ref{fig:resolution-overview}) are carried out together by reduction to a constrained optimization problem---in particular, MILP. 
A standard MILP problem involves finding values for a set of decision variables that maximize (or minimize) an objective function while satisfying a set of  constraints. We describe how our problem can be formulated into MILP\footnote{Without loss of generality, we show a formulation for a MILP problem to resolve a conflict between \emph{two} features.}. 

\subsubsection{Minimal requirements} As inputs, the resolver is given a pair of feature requirements in STL, $R_1$ and $R_2$. From these, the resolver first generates $weakSTL$ formulas for $R_1$ and $R_2$. In addition, the user specifies the maximum allowed degree of weakening by providing values for $p$ and $q$ for each \emph{weakSTL} requirement. These bounds, in effect, define the weakest possible requirement that is allowed by the \emph{weakSTL} requirement, also called the \emph{minimal requirement}. 

For example, consider $R_1 \equiv (alt(t) > 5)$. Suppose that the user is willing to accept $R_1$ to be weakened but no more than $(alt(t) > 2)$, since staying below that altitude is considered unacceptable. Then, the user would specify the value of $3$ for bound $p$ in \emph{weakSTL} $\varphi \equiv f_p(\signal{}(t)) > 0$, where $\varphi_0 \equiv R_1$.


\subsubsection{Conflict resolution}
Weakening-based resolution problem is formulated as follows:
\begin{problem}\label{prob:resolution}
Given (i) the original STL formulas $\varphi_0, \psi_0$ instantiated from $weakSTL~\varphi \text{ and } \psi$, respectively, (ii) past signal $\signal{}_{past}$ where $\neg\exists \signal{}_{pred} \bullet ({\signal{}_{past}} ^{\frown} \signal{}_{pred}, 0) \models \varphi_0 \land \psi_0$, find a pair of weakened STL formulas, $\varphi' \in M(\varphi), \psi' \in M(\psi)$, and a predictive signal, $\signal{}_{pred}$, such that 
{
\begin{align}
\exists\;\signal{}_{pred} \bullet ( {\signal{}_{past}} ^{\frown}  \signal{}_{pred}, 0) &\models \varphi' \land \psi' \label{prob1-conf-resolved}
\end{align}
}%
\end{problem}

Problem \ref{prob:resolution} defines a conflict as a condition where no possible future states (as represented by predictive signal $\signal{}_{pred}$) can satisfy the conjunction of the original requirements. Intuitively, weakening-based resolution attempts to resolve this conflict by finding  weakened requirements $\varphi'$ and $\psi'$ that are satisfiable over some possible future execution (i.e., ${\signal{}_{past}} ^{\frown}  \signal{}_{pred}$).


\subsubsection{Optimization problem} Problem \ref{prob:resolution} can be formulated as a constrained optimization problem, as follows: 

\begin{problem}\label{prob:optimization}
Given $\varphi,\psi \in weakSTL$, transition system $T = (\mathcal{Q},\mathcal{A},\delta,\mathcal{Q}_i)$, and state sequence $q^t = q_0,\dots,q_t$ representing the signal observed from the environment so far, compute: 

{\small
\vspace{-8pt}
\begin{align}
    \text{argmin}_{\theta(\varphi), \theta(\psi), \mathbf{a}} ~&~ \Delta(\varphi_0, \varphi_{\theta(\varphi)},\signal{},0) ~+~ \Delta(\psi_0, \psi_{\theta(\psi)}, \signal{},0) \label{eq:opt-objective} \\
    s.t. ~&~ \theta(\varphi) \in \mathbb{Z}^k     \label{eq:opt-phi}\\
    ~&~ \theta(\psi) \in \mathbb{Z}^m     \label{eq:opt-psi}\\
    ~&~ \signal{}_i = q_i\ \text{for } i\leq t  \label{eq:opt-past}\\
    ~&~ \signal{}_i = \delta(s_{i-1}, \mathbf{a}_{i-1})\ \text{for } t<i\leq t+N\label{eq:opt-future}\\ 
    ~&~ (\signal{}, 0)\models \varphi_{\theta(\varphi)} \land \psi_{\theta(\psi)} \label{eq:opt-sat}
\end{align}
}%
where $k$ and $m$ are the total number of weakening parameters of $\varphi$ and $\psi$ as defined by Eq.~\ref{eq:weakening-parameters}, $N\in \mathbb{N}$ is a finite horizon provide by the user, $\signal{} = s_0\dots s_{t+N}$, and $\mathbf{a} = a_t\dots a_{t+N-1}$. Moreover, transition system $T$ is provided by the environment model and is used to predict signal $\signal{}$ from $t$ to $t + N$.
\end{problem}

Problem~\ref{prob:optimization} poses the weakening-based resolution problem as an optimization problem.
Intuitively, this optimization problem generates (1) weakening parameters $\theta(\varphi), \theta(\psi)$ for weakSTL formulas $\varphi, \psi$ and (2) a sequence of $N$ actions $a_t\dots a_{t+N-1}$ that result in the satisfaction of the weakened requirements.
These parameters are generated such that the degree of weakening for each of the weakSTL formulas is minimized, i.e., weaken the requirements no more than necessary  (Eq. \ref{eq:opt-objective}).
Signal \signal{} is the concatenation of the past signal (Eq. \ref{eq:opt-past}) and the predicated signal generated via the environment model $T$ and actions $\mathbf{a}$ (Eq. \ref{eq:opt-future}). Finally,  the weakened requirements $\varphi_{\theta(\varphi)} \land \psi_{\theta(\psi)}$   must hold over \signal{} (Eq. \ref{eq:opt-sat}).



To encode Problem~\ref{prob:optimization} as an MILP instance, we extend the MILP-based formulation of the STL control synthesis problem in \cite{Raman2014} to capture the degree of weakening $\Delta(\varphi_0, \varphi_{\theta(\varphi)},\signal{},0)$ and $\Delta(\psi_0, \psi_{\theta(\psi)}, \signal{},0)$.
In \cite{Raman2014}, the control synthesis problem finds a signal $\signal{}$ that satisfies a given STL formula under an environmental model.
In our scenario, we include additional variables to capture the weakening parameters $\theta(\varphi)$ and $\theta(\psi)$ to the STL MILP encoding in \cite{Raman2014}.
As in \cite{Raman2014}, we assume that every predicate function $f$ that defines  $f(\signal{}(t))>0$ in STL formula $\varphi$ must be a linear or affine function. 
This assumption guarantees that our encoding is expressible as a MILP.

Finally, the translated MILP problem is dispatched to an off-the-shelf solver (Gurobi~\cite{gurobi} in our implementation). The solver can either find a solution or return UNSAT (meaning it cannot find weakened requirements and actions that satisfy the constraints). If the solver is able to find a solution, the resolver returns the generated action sequence $\mathbf{a} = a_t\dots a_{t+N-1}$ to be executed by the system as the resolved actions. 

\makeatletter
\newsavebox{\@brx}
\newcommand{\llangle}[1][]{\savebox{\@brx}{\(\m@th{#1\langle}\)}%
  \mathopen{\copy\@brx\kern-0.5\wd\@brx\usebox{\@brx}}}
\newcommand{\rrangle}[1][]{\savebox{\@brx}{\(\m@th{#1\rangle}\)}%
  \mathclose{\copy\@brx\kern-0.5\wd\@brx\usebox{\@brx}}}
\makeatother

\section{Implementation}
\label{sec:implementation}
\subsection{Simulator}
To demonstrate our approach, we have implemented a prototype of our resolution architecture\footnote{All of the code, models, and experimental data is available at \url{https://github.com/sychoo/CPS-weakening-based-resolution}} on top of PX4, an open source flight control software \cite{Lorenz:2015}.
To run our experiments, we use the jMAVSim drone simulator (part of PX4), which supports the simulation of the physical dynamics of the drone while it reacts to control actions. 
For evaluation, we implemented the following features on top of the PX4 flight control software:
\begin{itemize}
  \item \emph{Delivery Planning Feature}: Plans for the shortest path from point A to point B and generates a set of velocity vectors during the flight to follow the designated path. 
  \item \emph{Safe Landing Feature}: Directs the drone to land on the nearest land when the battery level drops below a preset safety threshold.
  \item \emph{Boundary Enforcer}: Ensures that the drone remains within the map boundaries. When active, it  generates a velocity vector orthogonal to the map boundary.
  \item \emph{Runaway Enforcer}: Ensures that the drone stays away from drones nearby. When active, it generates a velocity vector to evade a nearby drone.
\end{itemize}

\subsection{Environment Model}
As discussed in Section~\ref{sec:runtime-architecture}, our framework leverages a model of the environment during the resolution process. For the drone system, the environment model captures (1) 3D Cartesian space around the drone, (2) the physical dynamics of the drone, including how its location is changed by an action that sets its velocity, and how its speed is affected by an acceleration/deceleration action, (3)  the amount of remaining battery and its depletion rate (based on the velocity of the ego drone), and (4) the estimated speed and location of a nearby drone (which we call the \emph{chaser} drone); in particular, the model assumes that the chaser is moving towards the ego drone at a fixed speed. We believe that this model is general enough to capture important aspects of a typical drone environment and reusable across multiple features.

The environment model is specified in the MiniZinc modeling language and is automatically translated into the underlying MILP constraints during the resolution process.

\section{Evaluation}
\label{sec:evaluation}

This section presents the evaluation of our proposed approach. We focus on the following 3 research questions:
\begin{itemize}
 \item \textbf{RQ1}. Does the weakening-based approach better achieve the desired feature requirements compared to the existing priority-based approach? 
  \item 
  \textbf{RQ2}. Does the weakening-based approach provide a stronger guarantee for satisfying the minimal feature requirements compared to the existing priority-based approach? 
  \item 
  \textbf{RQ3}. What is the performance overhead of our approach? Does it interfere significantly with the system operation?
\end{itemize}

To study the proposed  questions, we conducted two case studies involving autonomous drones: (1) an organ delivery drone and (2) a surveillance drone. In each case study, we evaluated our weakening-based method to resolve conflicts between two different  features of the drone. 




\subsection{Experimental Design}
\label{sec:experimental-design}



We designed a set of experiments\footnote{All our experiments were run on a macOS machine with 32 GB RAM and a 6-core Intel Core i7.} to compare the proposed weakening-based approach to a priority-based resolution approach, which uses a fixed ordering among the features and only selects the action generated by the highest-ranking feature.
The weakening-based resolution approach does not require any ordering between different features. 
Instead, it allows the user to define an original requirement and a minimal requirement for each feature (specified as bounds on \emph{weakSTL} formulas, as described in Section~\ref{sec:milp}). 


When the MILP solver is unable to generate a solution, it outputs UNSAT (for unsatisfiability); this may occur in certain environmental contexts where it is impossible to weaken the requirements (within the maximum allowed degrees as defined by the minimal requirement). In that case, the resolver simply selects one of the conflicting actions to execute. 

To test the resolution approaches under diverse scenarios, we randomly generated different configurations for the drone simulation, including the initial starting points of the ego and nearby drones, the maximum speeds of the drones, the mission waypoints (e.g., delivery destination), and the size of the map boundary. Then, for each of these scenarios, we simulated the drone multiple times (each under a different resolution approach) and recorded the system states throughout its execution (i.e., the signal for the entire simulation). 

To measure the performance of the resolution approaches, we use the robustness of satisfaction of the given feature requirements as the metric. Our rationale behind choosing this metric is that  robustness captures  how well the system is achieving the  objectives of the features, and thus can serve as a reasonable proxy for the desirability of system behavior that results from a particular resolution approach. In particular, to analyze the impact of resolution, we measure and record the robustness values for the \emph{original} (not the weakened) requirements during the period of a feature interaction; that is, we start collecting the values starting at the time point where both features are activated and stop at the point where both features are deactivated.




Finally, before carrying out our experiments, we developed the following hypotheses to be tested:
\begin{itemize}
  \item \textbf{H1 (for RQ1).} The weakening-based  approach results in a higher overall satisfaction of feature requirements than the priority-based approach.
  \item \textbf{H2 (for RQ2).} The weakening-based approach provides a stronger guarantee of minimal requirements than the priority-based  approach.
  \item \textbf{H3 (for RQ3).} The weakening-based  approach incurs non-trivial overhead, but it is not significant enough to disrupt the operation of the existing drone controller.
\end{itemize}

\subsection{Organ Delivery Drone Case Study}
As  described in Section~\ref{sec:motivating-example}, there are two features of interest in the organ delivery drone: the \emph{delivery planner} and the \emph{the safe landing feature}.
A conflict between these two features arises when both features are activated simultaneously (i.e., the battery drops below a safe threshold while the planner is computing the next velocity vector  to the destination). 

The original STL requirements specified for the two features are as follows:

{\small
\vspace{-10pt}
\begin{align*}
  & R_{deliver}: \Box_{[0,1]}(curr\_speed > \\ 
  &\qquad\qquad\quad (distance\_to\_dest / remaining\_delivery\_time)) \\
  & R_{land}: \Box_{[0,1]}(battery < 40\% \rightarrow \Diamond_{[0,1]}(is\_landing = 1))
\end{align*}
}%
In addition, we also specified the following minimal requirements for the two features:

{\small
\vspace{-10pt}
\begin{align*}
  & R_{deliver}: \Box_{[0,1]}(curr\_speed > \\ 
  &\qquad\qquad\quad (distance\_to\_dest / remaining\_delivery\_time)) \\ 
  & R_{land}: \Box_{[0,1]}(battery < 20\% \rightarrow \Diamond_{[0,1]}(is\_landing = 1))
\end{align*}
}%
Note that the minimal requirement for the delivery planner is the same as the original one (i.e., it cannot be weakened), since timely delivery of the organ is considered critical.
 
For this case study, we generated 25 randomized  scenarios by varying the configuration parameters as described in Section~\ref{sec:experimental-design}. Then, we ran each scenario four times: (1) weakening-based approach with the delivery planner as the fallback action, (2) weakening with the landing feature as the fallback, (3) priority-based approach with the planner as the preferred feature, and (4) priority with the landing feature preferred. This resulted in a total of 100 scenario runs. 



\subsection{Surveillance Drone Case Study}

\begin{figure}[!t]
  \centering
  \includegraphics[width=0.50\linewidth]{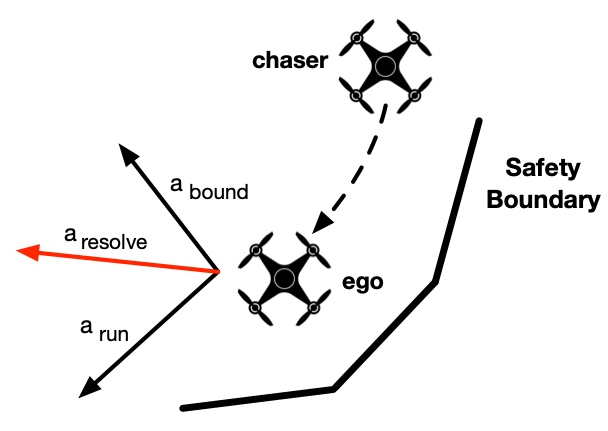}
  \caption{\small{A conflict scenario between the boundary and runaway enforcers. Action $a_{resolve}$ represents a possible action generated by the weakening-based approach.}}
  \vspace{-10pt}
  \label{fig:boundary-runaway-scenario}
\end{figure}

Consider a drone that performs a surveillance mission, visiting a set of waypoints within a designated boundary of the map (inspired by an example from \cite{Gafford2020}).
The environment  contains another simulated drone (called the \emph{chaser}) that constantly flies towards the ego drone. The two features of interest here are the \emph{boundary enforcer} and the \emph{runaway enforcer}, each of which is tasked with keeping the drone safe from a collision with the  boundary or the chaser (respectively). A conflict can occur in  situations when the ego drone travels to a position that is close to the boundary and the chaser, as shown in Figure~\ref{fig:boundary-runaway-scenario}.

The following original STL requirements were specified for the two features:

{\small
\vspace{-10pt}
\begin{align*}
  & R_{runaway}: \Box_{[0,1]}(distance\_to\_chaser > 10) \\
  & R_{boundary}: \Box_{[0,1]}(distance\_to\_boundary <= 20 \rightarrow \\ &\qquad\qquad\qquad\Diamond_{[0,1]}distance\_to\_boundary > 20)
\end{align*}
}%
In addition, we also specified the following minimal requirements for the two features:

{\small
\vspace{-10pt}
\begin{align*}
  & R_{runaway}: \Box_{[0,1]}(distance\_to\_chaser > 2) \\
  & R_{boundary}: \Box_{[0,1]}(distance\_to\_boundary <=20 \rightarrow \\ &\qquad\qquad\qquad\Diamond_{[0,1]}distance\_to\_boundary > 2)
\end{align*}
}%
As with the organ delivery case study, we  generated 25 random scenarios and ran each scenario four times (twice with the weakening-based approach and the other two times with the priority-based approach). 

\subsection{Experimental Result}

\begin{figure}[!t]
\includegraphics[width=\linewidth]{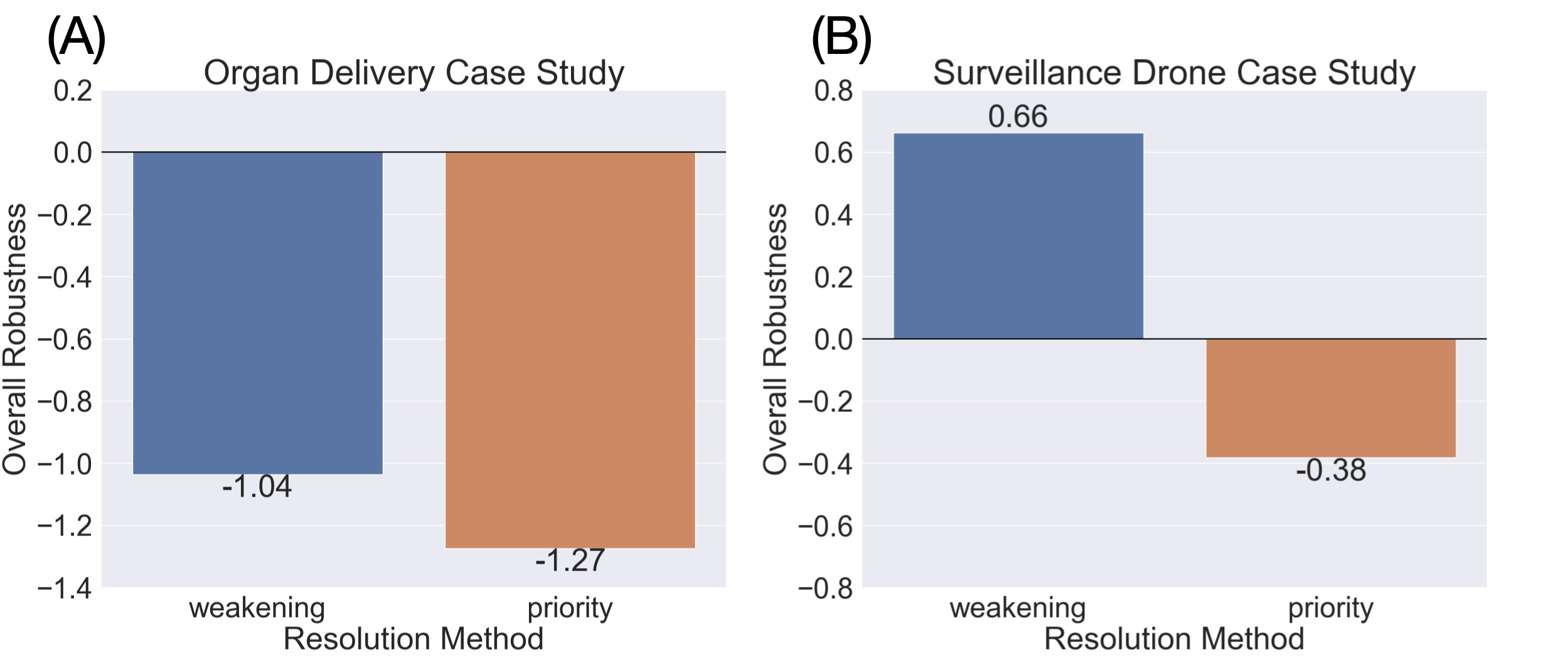}
  \caption{\small{Overall robustness values for the two case studies.}}
  \label{fig:overall-robustness}
  \vspace{-12pt}
\end{figure}

Figure~\ref{fig:overall-robustness} shows, for each case study, the \emph{overall} robustness values for the weakening-based and priority-based approaches. The overall robustness value is computed as the average of the normalized robustness values for the feature requirements and is intended to show how the system fulfills the objectives of all the features. As seen in Figure~\ref{fig:overall-robustness}, the weakening-based approach achieves higher overall robustness than the priority-based approach, as it attempts to satisfice the requirements of both features (unlike the priority-based method, which gives up on the feature that is not selected).

One may note that in the organ delivery case study, the overall robustness values for both approaches are negative; this is because during the conflicts in these scenarios, the actions available to the drone  are drastically different (land vs. keep flying) and thus selecting one feature will result in a large violation of the other's requirement. Even in such negative scenarios, the weakening-based approach is still able to achieve a lower overall violation of the requirements than the priority-based method, as shown in Figure~\ref{fig:overall-robustness}.

\begin{figure*}
  \includegraphics[width=\textwidth]{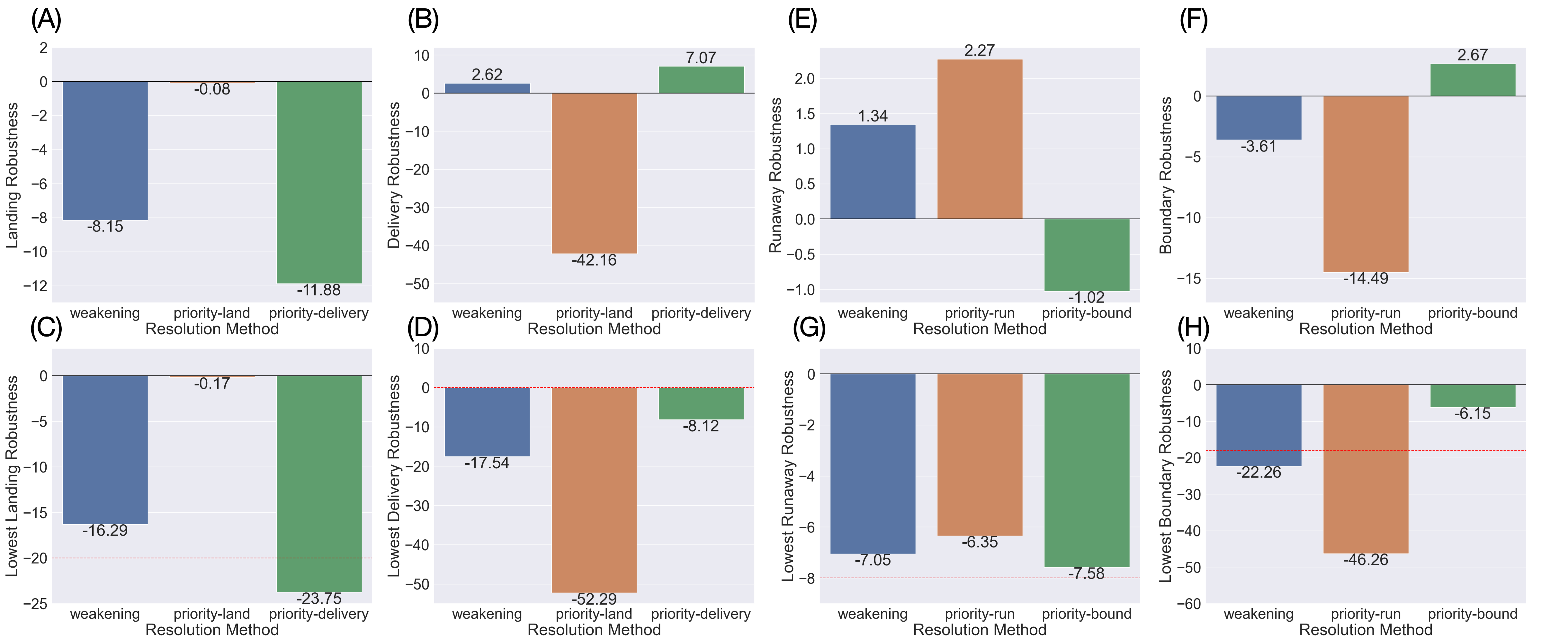}
    \caption{\small{Robustness breakdown by each feature requirement. Chart groups (A)-(D) and (E)-(H) correspond to the organ delivery and surveillance drone case studies, respectively. Each of (A), (B), (E), and (F) compares the average robustness values for three different approaches: (1) weakening-based, (2) priority-based with feature 1, and (3) priority with feature 2. Charts (C), (D), (G), and (H) show the lowest robustness values; the red line represents the threshold at which the minimal requirement is violated.}}
    \vspace{-10pt}
  \label{figure:feature-robustness}
\end{figure*}

We provide a more detailed analysis of the results, broken down by the individual feature requirements, shown in Fig.~\ref{figure:feature-robustness}.


\subsubsection{Organ delivery drone} 
In Fig.~\ref{figure:feature-robustness}, charts (A) and (B) show the average robustness values for the landing and delivery requirements, respectively. The results show that the priority-based approach achieves the highest robustness for the requirement of the feature that it selects (e.g., 7.07 for the delivery planning feature in (B)). At the same time, the requirement of the feature that is discarded by the priority-based method shows a large violation. Both of these outcomes are as expected, since the system achieves the requirement of the feature that it specifically prioritizes. 

In comparison, the results suggest that the weakening-based approach achieves a compromise between prioritizing one of the features and  discarding the other. For example, in chart (B), although the weakening-based approach achieves a lower robustness value than the priority method that selects the delivery feature, it avoids the large violation that would result if this feature was entirely discarded.

In (C), it can be seen that the weakening-based approach ensures the satisfaction of the minimal requirement (even in the worst case) while the priority-based method fails to do so. Lastly, in (D), none of the resolution methods can guarantee the delivery planning feature, as it is deemed as a hard constraint that cannot be weakened (i.e., the minimal requirement is the same as the original requirement). We did, however, observe that the weakening-based resolution still obtains a higher robustness value in its worst-case scenario than the priority-based method does.

\subsubsection{Surveillance drone} Similar patterns can be observed here as the ones in the prior case study. In Fig.~\ref{figure:feature-robustness}, charts (E) and (F), it can be seen that on average, the weakening-based approach achieves a compromise between prioritizing vs discarding a feature as it is done by the priority-based method. 

In (H), for the boundary enforcer, the weakening-based approach is not able to guarantee the minimal requirement, violating it in its worst-case outcome (-22.26). This is due to the fact that occasionally, the feature interaction scenario forces the drone into a non-recoverable position (i.e., cannot avoid crashing into the boundary), causing the resulting MILP problem to be unsatisfiable. Even then, it can be seen that the weakening-based approach avoids the large violation that is caused by the priority-based method (-46.26).

 \subsubsection{Summary} Based on Fig. \ref{fig:overall-robustness}, we conclude that the weakening-based approach attains a higher overall robustness value than the priority-based approach, supporting  hypothesis \textbf{H1}. This suggests that the weakening-based approach is effective at satisficing the requirements of both conflicting features.
 
 In Fig.~\ref{figure:feature-robustness}, (C-D), (G-H), it can be seen that the weakening-based approach, in its worst-case outcome, may  fail to guarantee the minimal requirement under environmental scenarios that do not permit any weakening solution. This suggests that if satisfying a particular requirement is critical, the priority-based method that always selects that feature may be more desirable. Thus, hypothesis \textbf{H2} is not supported.

\subsection{Performance Overhead}

Since the weakening-based approach uses a MILP solver, it incurs significantly more overhead than the priority-based method, which simply involves selecting the preferred action. Based on our timing measurement, the MILP-based resolution process took approximately 0.28 seconds on average across the two case studies. During our simulation runs, we did not observe any noticeable delays or disruptions to the drone operation. This is because the control loop inside the PX4 drone software was running at 2Hz, resulting in a window of 0.5 seconds for each cycle of the controller update (i.e., the resolution process completed before the next control action was to be generated).

On the other hand, the amount of overhead depends on the complexity of the feature requirements and the environment model, and thus it is possible that one may run into performance issues for larger, more complex systems than our drone software. As part of future work, we plan to explore other methods that leverage different types of search heuristics (e.g., machine learning-based or generic algorithms) and could potentially provide more efficient resolution.

\subsection{Threats to Validity}

There are three sources of potential errors in our experiments:
(1) the selected case study may not be representative of general CPS applications,
(2) the scenarios generated may exclude exceptional scenarios, and 
(3) drone simulation is hardware-dependent. 

For (1), we believe that the two case studies we conducted embody common characteristics of CPS, as PX4 is a well-established, popular drone software. However, a more extensive validation that involves other types of CPS (e.g., automotive systems) may provide further support for the effectiveness of the proposed resolution approach. 
To address (2), we generated a comprehensive sampling of scenarios and reduced selection biases by randomizing the configuration parameters. However, our sampling is non-exhaustive and is likely to exclude some exceptional scenarios.
For (3), more restrictive hardware may cause an increase in the performance overhead due to the computing resources that are required for the MILP solver and the simulator. 
\section{Related Work}
\label{sec:related-work}


There is a large body of work on feature interactions within software engineering ~\cite{fi-intro,nuseibeh-security,zave-fi,fi-detection-performance,feature-aware-verification,fi-detection-issre,fi-detection-strategies,fi-detection-jml,fi-detection-model-checking,atlee-fi-detection-fme, atlee-fse14,atlee-fse17,feature-detect1}. Here, we mainly provide a discussion of related work on  resolution (rather than detection) of feature interactions.

Gafford et al.~\cite{Gafford2020} proposes a \emph{synthesis-based} approach to the resolution of feature interactions, where given a pair of conflicting actions, a space of possible alternative actions is enumerated to find an action that best satisfies the objectives of the two features. Their approach is similar to ours in that it also (1) relies on the notion of robustness in STL to define the desirability of an action and (2) attempts to find an action as a middle-ground between the two conflicting actions. However, their approach is limited to cases where the features generate the same type of action (e.g., a pair of velocity vectors), whereas our approach can be applied to features with different types of actions (e.g., landing vs. flying towards the destination). 


Maia et al. investigates the problem of \emph{defiant components}, where one or more local components, in trying to achieve their individual objectives, conflict with a global system requirement~\cite{Maia2019}. They propose an approach called \emph{cautious adaption} to dynamically modify the behavior of the local component and fulfill the global requirement when a conflict arises; adaptation here is carried out by injecting a piece of logic
into the local component that overrides its default behavior.
Although their approach shares some similarities with ours in that they both involve temporarily changing the objective of a system component, there are some noticeable differences: (1) our work deals with conflicts between competing features (or components) instead of local vs. global requirements and (2) their approach requires wrappers that are  crafted at design time to handle known ``exceptional situations" (i.e., conflicts), whereas our approach can handle \emph{unexpected} conflict scenarios, as long as they are captured by the environment model.  


Requirement relaxation (or weakening) has been investigated in
self-adaptive systems. RELAX~\cite{Whittle2010} is a temporal logic to support the specification of requirements that explicitly capture uncertainty about possible system behavior. RELAX can support self-adaptation mechanisms where the system dynamically adjusts its behavior to accommodate for uncertainty or changes in the environment. DeVries et al. use RELAX to investigate the concept of \emph{partial} feature interactions, where a pair of features
only partially satisfy (i.e., satisfice) their individual requirements~\cite{DeVries19}, although they do not discuss a mechanism for resolving such interactions. One interesting future direction
is to leverage RELAX as another type of requirements specification language (instead of STL) to support the weakening-based resolution.

In \cite{Li2022}, the authors propose an iterative, \emph{multi-grained} approach to requirements relaxation,
where requirements of a higher granularity are relaxed first (for computational efficiency) before
lower-level requirements. Although our approach currently assumes feature requirements to be at the same level of granularity, their approach may be useful for resolving more complex
interactions that involve requirements across different levels of system abstraction.
 




Our approach for leveraging an environment model to generate an action that satisfies a desired objective
can be regarded as a type of model-predictive control (MPC)~\cite{mpc}. In particular, we adopt the STL-based MPC method developed by Raman et al.~\cite{raman-stl-synthesis}, where they also leverage a MILP solver to synthesize an action that satisfies an STL property.



\section{Limitations and Future Work}
\label{sec:limitations}


We propose a reconciliation-based approach to the resolution of feature interactions, where one or more of the given requirements are weakened to enable the conflicting features to behave consistently. Through case studies on autonomous drones, we have demonstrated that the proposed approach can achieve an overall higher satisfaction of the conflicting features, compared to the conventional priority-based method.


Our work makes several assumptions about the characteristics of the underlying system. First, it is most effective for systems where requirements can be assigned a meaningful, quantitative notion of satisfaction (i.e., STL-based requirements), are of equal importance, and where the user is willing to accept temporary degradation in the system performance of the original requirements (i.e., soft requirements). Thus, this approach may not be suited for features that perform a very critical function (e.g., emergency braking feature in a vehicle) where even small degradation is unacceptable; in such cases, a priority-based method may be more suitable. 


Our approach also relies on an environment model that generates predictive signals for different feature actions. Instead of a deterministic, discrete transition system, a stochastic model (e.g., Markov decision processes) may provide a more realistic model of the environment. Such a model would also require a very different approach to weakening than our MILP-based method and is beyond the scope of this paper. 



So far, we have evaluated our approach on pairwise feature interactions. Although, in principle, the weakening-based approach should be applicable to any number of features, a more extensive validation involving N-way interactions~\cite{n-way-interactions} would further support the generality of the proposed approach.



Finally, we believe that the idea of requirements weakening have other applications beside feature interaction resolution (e.g., using weakening to gracefully degrade the quality of a service in response to environmental deviations). We plan to explore such applications as part of future work. 



\section*{Acknowledgement}
 This material is based upon work funded and supported by the Department of Defense under Contract No.\ FA8702-15-D-0002 with Carnegie Mellon University for the operation of the Software Engineering Institute, a federally funded research and development center DM23-0069. This work was also supported in part by award N00014172899 from the Office of Naval Research, by the NSA under Award No. H9823018D0008, and by the National Science Foundation award CCF-2144860.

\bibliographystyle{IEEEtran}
\bibliography{seams22}

\begin{thebibliography}{10}
\providecommand{\url}[1]{#1}
\csname url@samestyle\endcsname
\providecommand{\newblock}{\relax}
\providecommand{\bibinfo}[2]{#2}
\providecommand{\BIBentrySTDinterwordspacing}{\spaceskip=0pt\relax}
\providecommand{\BIBentryALTinterwordstretchfactor}{4}
\providecommand{\BIBentryALTinterwordspacing}{\spaceskip=\fontdimen2\font plus
\BIBentryALTinterwordstretchfactor\fontdimen3\font minus
  \fontdimen4\font\relax}
\providecommand{\BIBforeignlanguage}[2]{{%
\expandafter\ifx\csname l@#1\endcsname\relax
\typeout{** WARNING: IEEEtran.bst: No hyphenation pattern has been}%
\typeout{** loaded for the language `#1'. Using the pattern for}%
\typeout{** the default language instead.}%
\else
\language=\csname l@#1\endcsname
\fi
#2}}
\providecommand{\BIBdecl}{\relax}
\BIBdecl

\bibitem{fi-intro}
M.~Calder, M.~Kolberg, E.~H. Magill, and S.~Reiff{-}Marganiec, ``Feature
  interaction: a critical review and considered forecast,'' \emph{Computer
  Networks}, vol.~41, no.~1, pp. 115--141, 2003.

\bibitem{nuseibeh-security}
A.~Nhlabatsi, R.~Laney, and B.~Nuseibeh, ``Feature interaction: The security
  threat from within software systems,'' \emph{Progress in Informatics},
  vol.~5, pp. 75--89, 2008.

\bibitem{zave-fi}
P.~Zave, ``Feature interactions and formal specifications in
  telecommunications,'' \emph{{IEEE} Computer}, vol.~26, no.~8, pp. 20--30,
  1993.

\bibitem{fi-iot}
L.~Yarosh and P.~Zave, ``Locked or not?: Mental models of {IoT} feature
  interaction,'' in \emph{Proceedings of the 2017 {CHI} Conference on Human
  Factors in Computing Systems, Denver, CO, USA, May 06-11, 2017.}, 2017, pp.
  2993--2997.

\bibitem{alma08}
A.~L.~J. Dominguez, N.~A. Day, and J.~J. Joyce, ``Modelling feature
  interactions in the automotive domain,'' in \emph{International Workshop on
  Modeling in Software Engineering (MiSE)}, 2008, pp. 45--50.

\bibitem{metzger04}
A.~Metzger, ``Feature interactions in embedded control systems,''
  \emph{Computer Networks}, vol.~45, no.~5, pp. 625--644, 2004.

\bibitem{atlee-fse17}
M.~H. Zibaeenejad, C.~Zhang, and J.~M. Atlee, ``Continuous variable-specific
  resolutions of feature interactions,'' in \emph{Proceedings of the 2017 11th
  Joint Meeting on Foundations of Software Engineering, {ESEC/FSE} 2017,
  Paderborn, Germany, September 4-8, 2017}, 2017, pp. 408--418.

\bibitem{lafortune-resolution}
Y.~Chen, S.~Lafortune, and F.~Lin, ``Resolving feature interactions using
  modular supervisory control with priorities,'' in \emph{Feature Interactions
  in Telecommunications Networks IV, June 17-19, 1997, Montr{\'{e}}al, Canada},
  1997, pp. 108--122.

\bibitem{atlee-resolution}
J.~D. Hay and J.~M. Atlee, ``Composing features and resolving interactions,''
  in \emph{{ACM} {SIGSOFT} Symposium on Foundations of Software Engineering, an
  Diego, California, USA, November 6-10, 2000, Proceedings}, 2000, pp.
  110--119.

\bibitem{atlee-ordering}
\BIBentryALTinterwordspacing
P.~A. Zimmer and J.~M. Atlee, ``Ordering features by category,'' \emph{Journal
  of Systems and Software}, vol.~85, no.~8, pp. 1782--1800, 2012. [Online].
  Available: \url{https://doi.org/10.1016/j.jss.2012.03.025}
\BIBentrySTDinterwordspacing

\bibitem{leung-ordering}
A.~Chavan, L.~Yang, K.~Ramachandran, and W.~H. Leung, ``Resolving feature
  interaction with precedence lists in the feature language extensions,'' in
  \emph{Feature Interactions in Software and Communication Systems IX,
  International Co nference on Feature Interactions in Software and
  Communication Systems, {ICFI} 2007, 3-5 September 2007, Grenoble, France},
  2007, pp. 114--128.

\bibitem{atlee-fse14}
C.~Bocovich and J.~M. Atlee, ``Variable-specific resolutions for feature
  interactions,'' in \emph{Proceedings of the 22nd {ACM} {SIGSOFT}
  International Symposium on Foundations of Software Engineering, (FSE-22),
  Hong Kong, China, November 16 - 22, 2014}, 2014, pp. 553--563.

\bibitem{rv18}
S.~G. Raghavan, K.~Watanabe, E.~Kang, C.~Lin, Z.~Jiang, and S.~Shiraishi,
  ``Property-driven runtime resolution of feature interactions,'' in
  \emph{Runtime Verification - 18th International Conference, {RV} 2018,
  Limassol, Cyprus, November 10-13, 2018, Proceedings}, 2018, pp. 316--333.

\bibitem{parnas-four-variable}
D.~L. Parnas and J.~Madey, ``Functional documents for computer systems,''
  \emph{Sci. Comput. Program.}, vol.~25, no.~1, pp. 41--61, 1995.

\bibitem{stl}
O.~Maler and D.~Nickovic, ``Monitoring temporal properties of continuous
  signals,'' in \emph{Formal Techniques, Modelling and Analysis of Timed and
  Fault-Tolerant Systems}.\hskip 1em plus 0.5em minus 0.4em\relax Springer
  Berlin Heidelberg, 2004, pp. 152--166.

\bibitem{Papadimitriou82}
C.~H. Papadimitriou and K.~Steiglitz, \emph{Combinatorial Optimization:
  Algorithms and Complexity}.\hskip 1em plus 0.5em minus 0.4em\relax
  Prentice-Hall, 1982.

\bibitem{px4}
{Dronecode Project}, ``{PX4 autopilot},'' \url{https://px4.io}, 2020.

\bibitem{sas-roadmap}
B.~H.~C. Cheng, R.~de~Lemos, H.~Giese, P.~Inverardi, J.~Magee, J.~Andersson,
  B.~Becker, N.~Bencomo, Y.~Brun, B.~Cukic, G.~D.~M. Serugendo, S.~Dustdar,
  A.~Finkelstein, C.~Gacek, K.~Geihs, V.~Grassi, G.~Karsai, H.~M. Kienle,
  J.~Kramer, M.~Litoiu, S.~Malek, R.~Mirandola, H.~A. M{\"{u}}ller, S.~Park,
  M.~Shaw, M.~Tichy, M.~Tivoli, D.~Weyns, and J.~Whittle, ``Software
  engineering for self-adaptive systems: {A} research roadmap,'' in
  \emph{Dagstuhl Seminar Report}, 2009, pp. 1--26.

\bibitem{Maia2019}
P.~H. Maia, L.~Vieira, M.~Chagas, Y.~Yu, A.~Zisman, and B.~Nuseibeh, ``Cautious
  adaptation of defiant components,'' in \emph{2019 34th IEEE/ACM International
  Conference on Automated Software Engineering (ASE)}, 2019, pp. 974--985.

\bibitem{simon1956rational}
H.~A. Simon, ``Rational choice and the structure of the environment,''
  \emph{Psychological Review}, vol.~63, no.~2, pp. 129--138, 1956.

\bibitem{ltl}
A.~Pnueli, ``The temporal logic of programs,'' in \emph{18th Annual Symposium
  on Foundations of Computer Science, Providence, Rhode Island, USA, 31 October
  - 1 November 1977}, 1977, pp. 46--57.

\bibitem{minizinc}
N.~Nethercote, P.~J. Stuckey, R.~Becket, S.~Brand, G.~J. Duck, and G.~Tack,
  ``Minizinc: Towards a standard cp modelling language,'' in \emph{Principles
  and Practice of Constraint Programming (CP)}.\hskip 1em plus 0.5em minus
  0.4em\relax Berlin, Heidelberg: Springer Berlin Heidelberg, 2007, pp.
  529--543.

\bibitem{Raman2014}
V.~Raman, A.~Donzé, M.~Maasoumy, R.~M. Murray, A.~Sangiovanni-Vincentelli, and
  S.~A. Seshia, ``Model predictive control with signal temporal logic
  specifications,'' in \emph{53rd IEEE Conference on Decision and Control},
  2014, pp. 81--87.

\bibitem{gurobi}
\BIBentryALTinterwordspacing
{Gurobi Optimization, LLC}, ``{Gurobi Optimizer Reference Manual},'' 2022.
  [Online]. Available: \url{https://www.gurobi.com}
\BIBentrySTDinterwordspacing

\bibitem{Lorenz:2015}
L.~Meier, D.~Honegger, and M.~Pollefeys, ``Px4: A node-based multithreaded open
  source robotics framework for deeply embedded platforms,'' in \emph{2015 IEEE
  International Conference on Robotics and Automation (ICRA)}, 2015, pp.
  6235--6240.

\bibitem{Gafford2020}
B.~Gafford, T.~Dürschmid, G.~A. Moreno, and E.~Kang, ``Synthesis-based
  resolution of feature interactions in cyber-physical systems,'' in \emph{2020
  35th IEEE/ACM International Conference on Automated Software Engineering
  (ASE)}, 2020, pp. 1090--1102.

\bibitem{fi-detection-performance}
N.~Siegmund, S.~S. Kolesnikov, C.~K{\"{a}}stner, S.~Apel, D.~S. Batory,
  M.~Rosenm{\"{u}}ller, and G.~Saake, ``Predicting performance via automated
  feature-interaction detection,'' in \emph{34th International Conference on
  Software Engineering, {ICSE} 2012, June 2-9, 2012, Zurich, Switzerland},
  2012, pp. 167--177.

\bibitem{feature-aware-verification}
S.~Apel, H.~Speidel, P.~Wendler, A.~von Rhein, and D.~Beyer, ``Detection of
  feature interactions using feature-aware verification,'' in \emph{26th
  {IEEE/ACM} International Conference on Automated Software Engineering {(ASE}
  2011), Lawrence, KS, USA, November 6-10, 2011}, 2011, pp. 372--375.

\bibitem{fi-detection-issre}
S.~Apel, W.~Scholz, C.~Lengauer, and C.~K{\"{a}}stner, ``Detecting dependences
  and interactions in feature-oriented design,'' in \emph{{IEEE} 21st
  International Symposium on Software Reliability Engineering, {ISSRE} 2010,
  San Jose, CA, USA, 1-4 November 2010}, 2010, pp. 161--170.

\bibitem{fi-detection-strategies}
S.~Apel, A.~von Rhein, P.~Wendler, A.~Gr{\"{o}}{\ss}linger, and D.~Beyer,
  ``Strategies for product-line verification: case studies and experiments,''
  in \emph{35th International Conference on Software Engineering, {ICSE} '13,
  San Francisco, CA, USA, May 18-26, 2013}, 2013, pp. 482--491.

\bibitem{fi-detection-jml}
W.~Scholz, T.~Th{\"{u}}m, S.~Apel, and C.~Lengauer, ``Automatic detection of
  feature interactions using the java modeling language: an experience
  report,'' in \emph{Software Product Lines - 15th International Conference,
  {SPLC} 2011, Munich, Germany, August 22-26, 2011. Workshop Proceedings
  (Volume 2)}, 2011, p.~7.

\bibitem{fi-detection-model-checking}
A.~Classen, P.~Heymans, P.~Schobbens, A.~Legay, and J.~Raskin, ``Model checking
  lots of systems: efficient verification of temporal properties in software
  product lines,'' in \emph{Proceedings of the 32nd {ACM/IEEE} International
  Conference on Software Engineering - Volume 1, {ICSE} 2010, Cape Town, South
  Africa, 1-8 May 2010}, 2010, pp. 335--344.

\bibitem{atlee-fi-detection-fme}
J.~M. Atlee, U.~Fahrenberg, and A.~Legay, ``Measuring behaviour interactions
  between product-line features,'' in \emph{3rd {IEEE/ACM} {FME} Workshop on
  Formal Methods in Software Engineering, FormaliSE 2015, Florence, Italy, May
  18, 2015}, 2015, pp. 20--25.

\bibitem{feature-detect1}
S.~Apel, A.~von Rhein, T.~Th{\"{u}}m, and C.~K{\"{a}}stner,
  ``Feature-interaction detection based on feature-based specifications,''
  \emph{Computer Networks}, vol.~57, no.~12, pp. 2399--2409, 2013.

\bibitem{Whittle2010}
J.~Whittle, P.~Sawyer, N.~Bencomo, B.~Cheng, and J.-M. Bruel, ``{RELAX}: A
  language to address uncertainty in self-adaptive systems requirement,''
  \emph{Requir. Eng.}, vol.~15, pp. 177--196, 06 2010.

\bibitem{DeVries19}
B.~DeVries and B.~H.~C. Cheng, ``Towards the detection of partial feature
  interactions,'' in \emph{Proceedings of the 14th International Symposium on
  Software Engineering for Adaptive and Self-Managing Systems {SEAMS}}.\hskip
  1em plus 0.5em minus 0.4em\relax {ACM}, 2019, pp. 146--152.

\bibitem{Li2022}
J.~Li and K.~Tei, ``Done is better than perfect: Iterative adaptation via
  multi-grained requirement relaxation,'' in \emph{IEEE International
  Conference on Requirements Engineering (RE)}, 2022.

\bibitem{mpc}
E.~F. Camacho and C.~B. Alba, \emph{Model predictive control}.\hskip 1em plus
  0.5em minus 0.4em\relax Springer science \& business media, 2013.

\bibitem{raman-stl-synthesis}
V.~Raman, A.~Donz{\'{e}}, M.~Maasoumy, R.~M. Murray, A.~L.
  Sangiovanni{-}Vincentelli, and S.~A. Seshia, ``Model predictive control with
  signal temporal logic specifications,'' in \emph{53rd {IEEE} Conference on
  Decision and Control, {CDC} 2014, Los Angeles, CA, USA, December 15-17,
  2014}, 2014, pp. 81--87.

\bibitem{n-way-interactions}
B.~DeVries and B.~H.~C. Cheng, ``Run-time monitoring of self-adaptive systems
  to detect n-way feature interactions and their causes,'' in \emph{Proceedings
  of the 13th International Conference on Software Engineering for Adaptive and
  Self-Managing Systems (SEAMS)}, 2018, pp. 94--100.

\end{thebibliography}

\end{document}